\documentclass{article}
\usepackage{amssymb}
\usepackage{amsmath}
\usepackage[dvips]{graphicx}
\usepackage{amsfonts}

\input{tcilatex}

\begin{document}

\title{Tunnelling From G\"{o}del Black Holes}
\author{Ryan Kerner\thanks{%
rkerner@sciborg.uwaterloo.ca} and R.B. Mann\thanks{%
rbmann@sciborg.uwaterloo.ca} \\
Department of Physics \&\ Astronomy, University of Waterloo\\
Waterloo, Ontario N2L 3G1, Canada}
\maketitle

\begin{abstract}
We consider the spacetime structure of Kerr-G\"{o}del black holes, analyzing
their parameter space Kerr-G\"{o}del in detail. \ We apply the tunnelling
method to compute their temperature and compare the results to previous
calculations obtained via other methods. \ We claim that it is not possible
to have the CTC horizon in between the two black hole horizons and include a
discussion of issues that occur when the radius of the CTC horizon is
smaller than the radius of both black hole horizons. \ 
\end{abstract}

\section{Introduction}

There has been a fair amount of activity in recent years studying G\"{o}%
del-type solutions to 5d supergravity\textbf{\ }\cite{5d godel}-\cite%
{general thermo}. Various black holes embedded in G\"{o}del universe
backgrounds have been obtained as exact solutions\ \cite{godel-Reissner
black hole,gimonhash,rotation in godel} and their string-theoretic
implications make them a lively subject of interest. For example G\"{o}del
type solutions have been shown to be T-dual to pp-waves \cite{godel dual to
p waves}-\cite{Harmark}. \ Since closed-timelike curves (CTCs) exist in G%
\"{o}del spacetimes these solutions can be used to investigate the
implications of CTCs for string theory \cite{ctcs1,ctcs2,ctcs3,ctcs4}. \ 

\textbf{\ }The black hole solutions are of the Schwarzschild-Kerr type
embedded in a G\"{o}del universe \cite{gimonhash}. A study of their
thermodynamic behaviour \cite{shwarz-godel thermo,general thermo} has
indicated that the expected relations of black hole thermodynamics are
satisfied. Making use of standard Wick-rotation methods,\ their temperature
has been shown to equal $\kappa /2\pi $\ (where $\kappa $\ is the surface
gravity) their entropy to equal $A/4$\ (where $A$\ is the surface area of
the black hole) and the first law of thermodynamics to be satisfied.\textbf{%
\ \ }

We consider in this paper an analysis of the Kerr-G\"{o}del\ spacetime,
employing the tunnelling method \cite{early tunnelling}-\cite{recent2} to
analyze its thermodynamic properties. The tunnelling method is a
semi-classical approach to black hole radiation that allows one to calculate
the temperature in a manner independent of the traditional Wick Rotation
methods of temperature calculation. As such it provides a useful cross-check
on the thermodynamic properties of these objects and has been shown to be
quite robust, having been applied to a variety of different spacetimes such
as the Kerr and Kerr-Newmann\ cases\cite{kerr and kerr newman, Zhang and
Zhao, Last paper}, black rings \cite{Black Rings}, the 3-dimensional AdS
black hole \cite{Vargenas, BTZ}, and the Vaidya \cite{Vaidya}, and Taub-NUT
spacetimes \cite{Last paper}\textbf{. \ } The presence of CTCs merits
consideration of the applicability of the tunnelling method to Kerr-G\"{o}%
del spacetimes. Due to the presence of a CTC horizon (in addition to the
usual black-hole horizons)\ some qualitatively new features appear. Our
investigation of these spacetimes is in large part motivated by the fact
that these new features provide additional tests as to the robustness of the
tunnelling approach.

\ We being by reviewing the Kerr-G\"{o}del spacetime and some of its
properties. We then describe, in section 3, properties of its parameter
space and show that either the CTC horizon is outside both black hole
horizons, inside both black hole horizons, or in coincidence with one of the
horizons. We claim that it is not possible for the CTC horizon to be
strictly in between the two black hole horizons, a property previously
overlooked in discussions of this spacetime \cite{general thermo}. \ \textbf{%
\ }We then quickly review the tunnelling method and apply it to calculate
the temperature of Kerr-G\"{o}del spacetimes, showing consistency with
previous results. We extend our investigation further insofar as we\ include
a brief discussion of the issues that occur when the CTC horizon is inside
the black hole horizons.\textbf{\ }

\section{Review of 5d Kerr-G\"{o}del Spacetimes}

The 5d Kerr-G\"{o}del spacetime has the metric \cite{gimonhash}\textbf{\ }%
\begin{eqnarray}
ds^{2} &=&-f(r)(dt+\frac{a(r)}{f(r)}\sigma _{3})^{2}+\frac{dr^{2}}{V(r)}+%
\frac{r^{2}}{4}(\sigma _{1}^{2}+\sigma _{2}^{2})+\frac{r^{2}V(r)}{4f(r)}%
\sigma _{3}^{2}  \label{metric 1st form} \\
A &=&\frac{\sqrt{3}}{2}jr^{2}\sigma _{3}  \label{gauge}
\end{eqnarray}%
where%
\begin{eqnarray*}
f(r) &=&1-\frac{2m}{r^{2}} \\
a(r) &=&jr^{2}+\frac{ml}{r^{2}} \\
V(r) &=&1-\frac{2m}{r^{2}}+\frac{16j^{2}m^{2}}{r^{2}}+\frac{8jml}{r^{2}}+%
\frac{2ml^{2}}{r^{4}}
\end{eqnarray*}%
and the $\sigma $'s are the right-invariant one-forms on SU(2), with Euler
angles ($\theta ,\phi ,\psi $): 
\begin{eqnarray*}
\sigma _{1} &=&\sin \phi d\theta -\cos \phi \sin \theta d\psi \\
\sigma _{2} &=&\cos \phi d\theta +\sin \phi \sin \theta d\psi \\
\sigma _{3} &=&d\phi +\cos \theta d\psi
\end{eqnarray*}%
This metric may be obtained by embedding the Kerr black hole metric (with
the two possible rotation parameters set to the same value i.e. $%
l_{1}=l_{2}=l$) in a 5-d G\"{o}del universe.

This metric and gauge field satisfy the following 4+1 dimensional equations
of motion:%
\begin{equation*}
R_{\mu \nu }=2(F_{\mu \alpha }F_{\nu }^{\text{ \ }\alpha }-\frac{1}{6}g_{\mu
\nu }F^{2}),\text{ }D_{\mu }F^{\mu \nu }=\frac{1}{2\sqrt{3}}\tilde{\epsilon}%
^{\alpha \beta \gamma \mu \nu }F_{\alpha \beta }F_{\gamma \mu }
\end{equation*}%
where 
\begin{equation*}
\tilde{\epsilon}_{\alpha \beta \gamma \mu \nu }=\sqrt{-g}\epsilon _{\alpha
\beta \gamma \mu \nu }
\end{equation*}%
Some other useful ways to write the metric (\ref{metric 1st form}) are the
expanded form: 
\begin{equation*}
ds^{2}=-f(r)dt^{2}-2a(r)dt\sigma _{3}+g(r)\sigma _{3}^{2}+\frac{dr^{2}}{V(r)}%
+\frac{r^{2}}{4}(\sigma _{1}^{2}+\sigma _{2}^{2})
\end{equation*}%
where 
\begin{equation}
g(r)=\frac{r^{2}V(r)-4a^{2}(r)}{4f(r)}=-j^{2}r^{4}+\frac{1-8mj^{2}}{4}r^{2}+%
\frac{ml^{2}}{2r^{2}}  \label{gr-def}
\end{equation}%
and the lapse-shift form:%
\begin{equation}
ds^{2}=-N^{2}dt^{2}+g(r)(\sigma _{3}-\frac{a(r)}{g(r)}dt)^{2}+\frac{dr^{2}}{%
V(r)}+\frac{r^{2}}{4}(\sigma _{1}^{2}+\sigma _{2}^{2})  \label{LSform}
\end{equation}%
where 
\begin{equation*}
N^{2}=f(r)+\frac{a^{2}(r)}{g(r)}=\frac{r^{2}V(r)}{4g(r)}
\end{equation*}

When the parameters $j$ and $l$ are set to zero the metric simply reduces to
the 5d Schwarzschild black hole, whose mass is proportional to the parameter 
$m$. The parameter $j$ is the G\"{o}del parameter and is responsible for the
rotation of the spacetime; when $m=l=0$ the metric reduces to that of the
5-d G\"{o}del universe \cite{5d godel}. The parameter $l$ is related to the
rotation of the black hole. \ When $j=0$ this reduces to the 5d Kerr black
hole with the two possible rotation parameters ($l_{1},l_{2}$) of the
general 5-d Kerr spacetime set equal to $l$. \ When $l=0$ the solution
becomes the Schwarzschild-G\"{o}del black hole. \ The metric is well behaved
at the horizons and the scalars only become singular at the origin. \ It has
recently been noted that\ the gauge field is not well behaved at the
horizons \cite{general thermo} although it is possible to pass to a new
gauge potential that is well behaved. When $g(r)<0$ then $\partial _{\phi }$
will be timelike, indicating the presence of closed timelike curves since $%
\phi $ is periodic. The point at which $g(r)=0$\ is where the lapse ($N^{2}$%
) becomes infinite, implying that nothing can cross over to the CTC region
from the region without CTC's. \ This property is implied by the geodesic
solutions for Schwarzschild-G\"{o}del found in \cite{gimonhash} but we will
argue later in the paper that this is a general property of Kerr-G\"{o}del.\
The lapse vanishes when $V(r)=0$; these points correspond to the black hole
horizons.

The function $f(r)$ is equal to zero when $r=\sqrt{2m}$, corresponding to an
ergosphere. The angular velocity of locally nonrotating observers is given
by $\Omega =\frac{d\phi }{dt}=\frac{a(r)}{g(r)}$ with $\Omega _{H}=\frac{%
a(r_{H})}{g(r_{H})}$ denoting the angular velocity of the horizon. There is
a special choice of parameters that will cause the angular velocity at the
horizon to vanish (besides the trivial $l=j=0$). When $l=-4jm$ then $V(r)=0$
has solutions at $r^{2}=2m$ and $r^{2}=16j^{2}m^{2}$. The function $a(r)$
will be equal to zero for $r^{2}=2m$. Consequently $\Omega _{H}$ will vanish
for the choice $l=-4jm$ at the horizon $r=\sqrt{2m}.$

For the case $l$ $=0$ there is only one black hole horizon located at $r_{H}=%
\sqrt{2m(1-8j^{2}m)}$. \textbf{\ }Clearly $1>8j^{2}m$ for the horizon to be
well defined. A standard Wick-rotation approach yields a temperature $T_{H}=%
\frac{1}{2\pi \sqrt{2m(1-8j^{2}m)^{3}}}$\ for the Schwarzschild-G\"{o}del
black hole \cite{shwarz-godel thermo}, where the horizon has angular
velocity $\Omega _{H}=\frac{4j}{(1-8j^{2}m)^{2}}$. There will be no CTC's
for $r<r_{CTC}=\frac{\sqrt{(1-8mj^{2})}}{2j}$ (the region where $g(r)>0$),
and the condition $r_{CTC}>r_{H}$ corresponds to $1>8j^{2}m$. Hence for $l=0$
the CTC horizon is always outside of the black hole horizon. This property
is not true for $l\neq 0$ and in the next section we will investigate the
conditions under which the CTC horizon is no longer outside of the black
hole horizons.

\section{\ Analysis of 5d Kerr-G\"{o}del}

\subsection{Parameter Space of the 5d Kerr-G\"{o}del}

We will start by examining the parameter space of the 5d Kerr-G\"{o}del
spacetimes. \ The functions of interest are $g(r)=0$, which determines the
location of the CTC horizon and $V(r)=0$ which determines the black hole
horizons. \ We wish to find out how the horizons behave in terms of the
parameters $l$\ and $j$. \ To simplify the analysis we will reparameterize
as follows%
\begin{equation}
J=j\sqrt{8m}\qquad L=\frac{l}{\sqrt{2m}}\qquad x=\frac{r^{2}}{2m}
\label{repam}
\end{equation}%
so $x=1$ at the ergosphere ($r^{2}=2m$), $J^{2}=1$ when $8mj^{2}=1$, and the
special case $l=-4jm$ corresponds to the choice $L=-J$.

The equations $V(r)=0$ and $g(r)=0$ now correspond respectively to the
equations:%
\begin{eqnarray}
\frac{1}{x^{2}}(x^{2}-(1-J^{2}-2LJ)x+L^{2}) &=&0  \label{horizons} \\
-\frac{m}{2x}(J^{2}x^{3}-(1-J^{2})x^{2}-L^{2}) &=&0  \label{ctc}
\end{eqnarray}%
There are two solutions to the quadratic equation (\ref{horizons}) and there
is only one real solution to (\ref{ctc}) \ when $L$\ is non-zero (it can be
shown that when $L=0$\ only the single non-zero solution of (\ref{ctc}) is
relevant). \ \ The solutions of (\ref{horizons}) and (\ref{ctc}) are
respectively: 
\begin{eqnarray}
x_{\pm } &=&\frac{1}{2}\left( (1-J^{2}-2LJ)\pm \sqrt{(1-J^{2}-2LJ)^{2}-4L^{2}%
}\right)  \label{xH} \\
x_{ctc} &=&\frac{C(J,L)}{6J^{2}}+\frac{2(1-J^{2})^{2}}{3J^{2}C(J,L)}+\frac{%
1-J^{2}}{3J^{2}}  \label{xctc}
\end{eqnarray}%
where 
\begin{equation*}
C(J,L)=\left[ 108L^{2}J^{4}+8(1-J^{2})^{3}+12J^{2}\sqrt{%
3L^{2}(27L^{2}J^{4}+4(1-J^{2})^{3}}\right] ^{\frac{1}{3}}
\end{equation*}

\begin{figure}[tbp]
\centering
\includegraphics[
height=3.1687in,
width=3.9721in
]{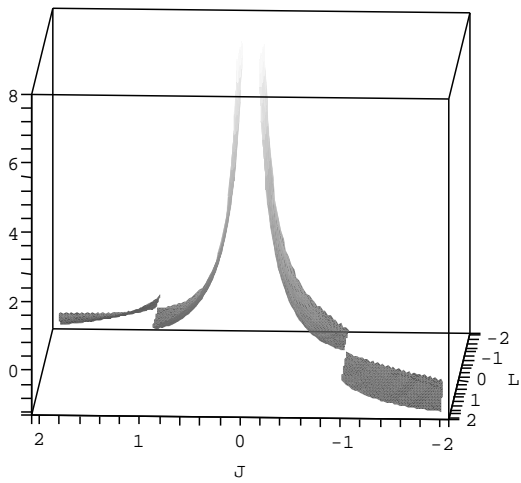}
\caption{3d plot of $\protect\sqrt{x_{ctc}}-\protect\sqrt{x_{+}}$ in terms
of $L$ and $J$.(i.e. compares location of the CTC horzion to largest black
hole horizon) \ Note: Regions when $J^{2}>1$ are negative which means the
CTC horizon is inside the black hole horizon. \ The peak at $J=0$
corresponds to the infinite ctc horizon and indicates regular 5d Kerr.}
\label{figure1}
\end{figure}

\begin{figure}[tbp]
\centering
\includegraphics[
height=3.1687in,
width=3.1687in
]{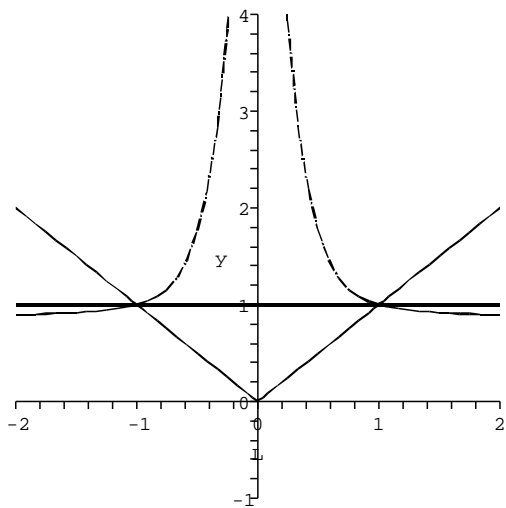}
\caption{Plots of $\protect\sqrt{x_{ctc}},\protect\sqrt{x_{-}},\protect\sqrt{%
x_{+}}$ it terms of $L$ and when $J=-L$, Note: dashed line corresponds to $%
\protect\sqrt{x_{ctc}}$ and solid lines are $\protect\sqrt{x_{-}},\protect%
\sqrt{x_{+}}$. \ Notice the CTC is horizon is either outside both horizons
or inside both horizons but never in between.}
\label{Figure1a}
\end{figure}

\begin{figure}[tbp]
\centering
\includegraphics[
height=1.7573in,
width=3.9366in
]{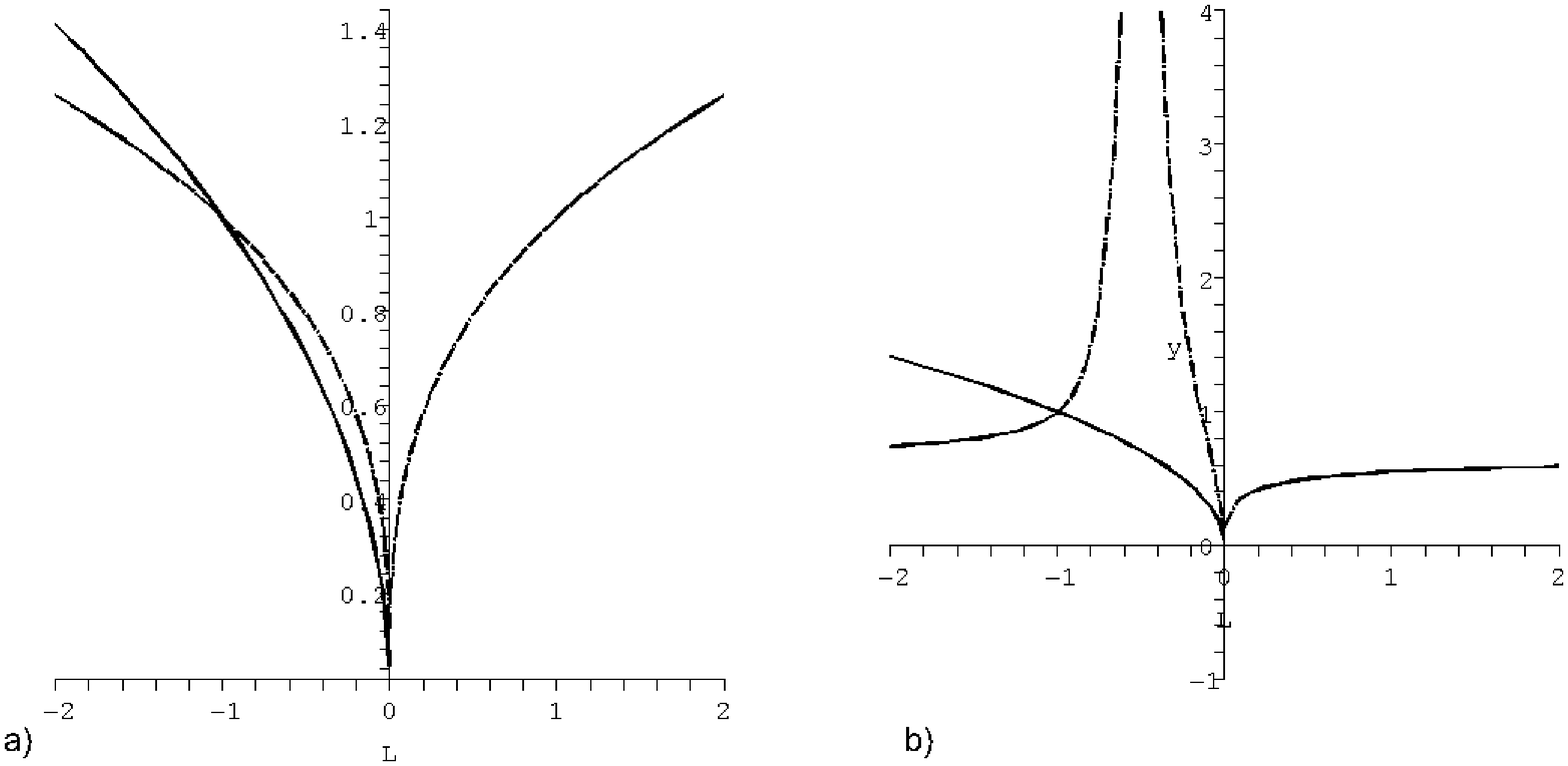}
\caption{Plots of $\protect\sqrt{x_{ctc}},\protect\sqrt{x_{-}},\protect\sqrt{%
x_{+}}$ it terms of a) $L$ when $J=1$ b) $L$ \ when $J=-2L-1$\ Note: the
dashed line corresponds to $\protect\sqrt{x_{ctc}}$ and solid lines are $%
\protect\sqrt{x_{-}},\protect\sqrt{x_{+}}$ (These are extremal cases so $%
\protect\sqrt{x_{-}}=\protect\sqrt{x_{+}}$ ) Notice, black hole horizons do
not exist for a) when $L>0$ and b) when $L>0$. \ Also in b) $\protect\sqrt{%
x_{ctc}}$ is infinite at $L=-\frac{1}{2}$ because. $J=0$ which is 5d Kerr.}
\label{fig1bandc}
\end{figure}

The black hole is extremal when $x_{+}=x_{-}$ and will occur when $J=\pm
1,J=-2L+1,J=-2L-1$. \ All three horizons will coincide when $J=-L=\pm 1.$ \
Note that black hole horizons only exist when $x_{\pm }>0$ are since the
horizon radii $r_{\pm }=\sqrt{2mx_{\pm }}$.

\ In Figure \ref{figure1} we show a 3d plot of $\sqrt{x_{ctc}}-\sqrt{x_{+}}$%
\ in \ terms of $L$\ and $J$. \ Note that when $J^{2}>1$\ the value of $%
\sqrt{x_{ctc}}-\sqrt{x_{+}}$\ is negative so the CTC horizon is inside the
black hole horizon. \textit{\ }In order to get a feel for how the horizons
behave it is useful to plot all three horizons (inner, outer, and CTC)
together for special values of $J$. \ The choices of $J$\ that are
interesting are $J=-L$\ (which is when $\Omega _{H}=0$\ at the horizon
located at $x=1$), and the extremal cases\ $J=1$\ and $J=-2L-1$. \textit{\ }%
These plots are shown in Figures \ref{Figure1a}, \ref{fig1bandc} a) and b)
respectively. \ \ Notice that for figure \ref{Figure1a} the CTC horizon is
either outside both of $r_{+}$ and $r_{-}$ (ie the black hole horizons) or
inside both $r_{+}$ and $r_{-}$ \ (this is also trivially true for the other
two plots since they are extremal black holes). \ In all three plots the
change from CTCs outside the black hole horizons to inside the horizons
occurs when you go beyond the points $J=-L=\pm 1$.

\begin{figure}[tbp]
\centering
\includegraphics[
height=3.1678in,
width=3.1678in
]{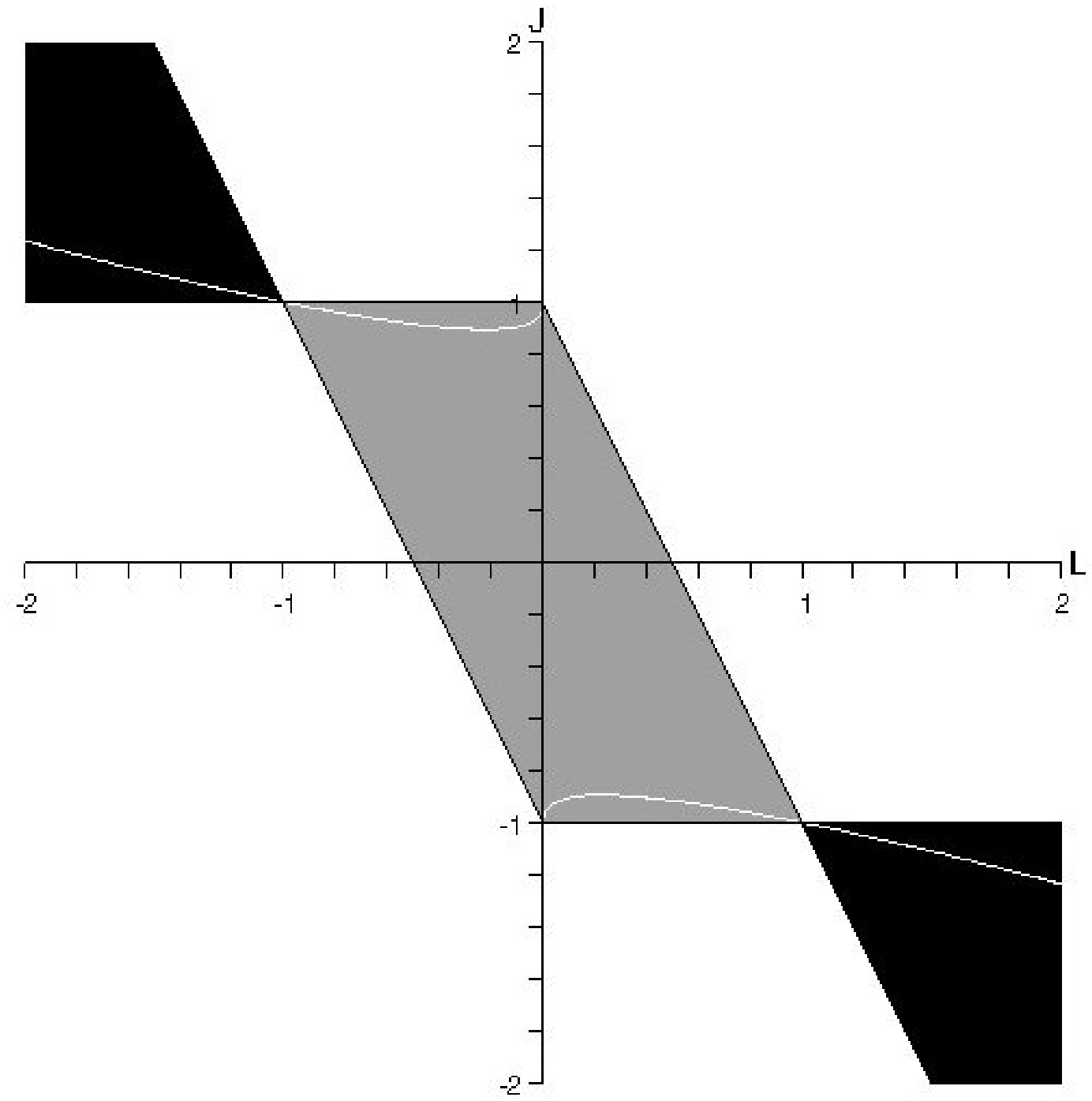}
\caption{Plot of the horizon behavior in terms of $J$ and $L$. \ The white
region corresponds to naked singularities (no black hole horizons). \ In the
grey region the CTC\ horizon is outside both black hole horizons. \ In the
black region the CTC horizon is inside both black hole horizons. \ The white
line corresponds to the special case when the CTC horizon is in coincidence
with a black hole horizon (outer horizon in grey region, inner horizon in
black region, and both at the special points $J=-L=\pm 1$)}
\label{figure2}
\end{figure}

We claim that is not possible to have the CTC\ horizon located in between
the two black hole horizons. Assuming the contrary, consider the problem of
finding values of $J$\ and $L$\ when the CTC horizon is in between the two
black hole horizons. \ We first look for solutions when the CTC horizon is
in coincidence with one of the black hole horizons. \ We find that $%
x_{ctc}=x_{\pm }$\ when the equation\textbf{\ }%
\begin{equation}
(3J^{2}+2JL-2)^{2}+4J^{2}-5J^{4}=0  \label{overlap}
\end{equation}%
holds. Notice that\ $J=-L=\pm 1$\ are solutions to (\ref{overlap}).

An analysis of the curve resulting from the left-hand-side of (\ref{overlap}%
) indicates that when both $J^{2}<1$\ and $L^{2}<1$\ then the CTC horizon is
coincident with the outer horizon; on either side of this curve the CTC
horizon is outside both black hole horizons. When both $J^{2}>1$\ and $%
L^{2}>1$\ then the CTC horizon is coincident with the inner horizon, and on
either side of this curve the CTC horizon is inside both $r_{+}$ and $r_{-}$%
. \ \ In all other regions of \ \ parameter space the metric (\ref{LSform})
has naked singularities. Figure \ref{figure2} \ illustrates this behavior in
terms of $J$\ and $L$. \ In the grey region the CTC horizon is outside both
black hole horizons. \ In the black region the CTC horizon is inside both $%
r_{+}$ and $r_{-}$. \ The white line corresponds to the curves resulting
from (\ref{overlap}). In the white region the metric has\ no black hole
horizons and naked singularities are present.\ \ 

An alternate verification for the fact that the CTC horizon is never in
between the black hole horizons may be obtained by substituting $x_{\pm }$\
into $g(x)$, which shows that when $J^{2}<1$\ then $g(x_{-})>0,g(x_{+})\geq
0 $\ and for $J^{2}>$\ $1$\ then $g(x_{-})\leq 0,g(x_{+})<0$\ (plots not
shown).\textit{\ \ }Conceptually it is easy to see why this property must be
true by looking at the definition (\ref{gr-def}) of the function $g(r)$,
which defines where the CTC horizon must be located. \ If $r=r_{ctc}$ then $%
g(r_{ctc})=0$ which implies $r_{ctc}^{2}V(r_{ctc})=4a^{2}(r_{ctc})$. \ For
this equality to be true then $V(r_{ctc})$\ must be positive since every
other term in the equation is positive. \ Since $V(r_{ctc})$ cannot be
negative then $r_{ctc}$\ cannot be in between $r_{-}$\ and $r_{+}$.

Another property worth mentioning is the location of the black hole horizons
in relation to $x=1$ (ie. the ergosphere $r=\sqrt{2m}$). \ When $J^{2}<1$
then $x_{\pm }\leq 1$ so the horizons are inside the ergosphere. When $%
J^{2}>1$ then $x_{\pm }\geq 1$ so the \textquotedblleft
horizons\textquotedblright\ are outside the ergosphere. Indeed\ when $%
J^{2}>1 $\ the surfaces $x_{\pm }=1$\ are not actually horizons, though we
have been using this term as a counterpart to the $J^{2}<1$\ case.
Henceforth we shall refer to this as the \textquotedblleft other
region\textquotedblright\ of parameter space.

Our finding that the CTC horizon can never be in between the two black hole
horizons is contrary to assumptions made in previous work \cite{general
thermo}. However the resultant thermodynamics is not significantly altered,
as all main results consider only the situation when the CTC horizon is
outside the black hole. \ In the next two sections we will discuss the
properties of the black hole region and the other region of parameter space. 
\textbf{\ }\textit{\ }

\subsection{Black Hole region of parameter space ($J^{2}<1$)}

This is the region that is well understood and can be simply regarded as a
Kerr Black Hole embedded in a G\"{o}del space time, with the CTC horizon
outside of the black hole horizons. To better understand this case we will
take a look at the geodesics in the $(t,r,\phi )$ plane (with $\theta $ and $%
\psi $ fixed). The metric becomes 
\begin{equation}
ds^{2}=-\frac{r^{2}V(r)}{4g(r)}dt^{2}+g(r)(d\phi -\frac{a(r)}{g(r)}dt)^{2}+%
\frac{dr^{2}}{V(r)}  \label{reduce lapse}
\end{equation}%
Note that $g(r_{H})\geq 0$ for the choice of parameters $(-1\leq J\leq 1,-%
\frac{1}{2}-\frac{J}{2}\leq L\leq \frac{1}{2}-\frac{J}{2})$ that we are
considering. For convenience we impose the further restriction that $L\neq 
\frac{-3J^{2}+2+\sqrt{5J^{4}-4J^{2}}}{2J}$ and $L\neq \frac{-3J^{2}+2-\sqrt{%
5J^{4}-4J^{2}}}{2J}$ so that $g(r_{H})>0$ and the CTC horizon is strictly
outside the outer black hole horizon.

The tangent vector to a geodesic is given by:%
\begin{equation*}
u^{\alpha }=[\dot{t},\dot{r},\dot{\phi}]
\end{equation*}%
where dot denotes the derivative with respect to the affine parameter $%
\lambda $. \ For this metric $\partial _{t}$ and $\partial _{\phi }$ are
Killing vectors so in general the energy and angular momentum for these
geodesics are respectively 
\begin{eqnarray*}
E &=&\frac{r^{2}V(r)-4a^{2}(r)}{4g(r)}\dot{t}+a(r)\dot{\phi} \\
\ell &=&-a(r)\dot{t}+g(r)\dot{\phi}
\end{eqnarray*}%
We are interested in geodesics with $\ell =0$. \textit{\ }Note that for
constant $r$\ the quantity $d\chi =d\phi -\frac{a(r)}{g(r)}dt$\ is constant
(i.e. $\frac{d\chi }{d\lambda }=0$); for $r=r_{H}$\ these correspond to
geodesics for which $\chi =\phi -\Omega _{H}t$ is constant on the horizon
(recall $\Omega _{H}=\frac{a(r_{H})}{g(r_{H})}$). \ 

Setting $\ell =0$ yields $\frac{a(r)}{g(r)}$ $\dot{t}=\dot{\phi}$ and $E=%
\frac{r^{2}V(r)}{4g(r)}\dot{t}$. For null geodesics we find $\dot{r}=\pm 
\frac{rV(r)}{2\sqrt{g(r)}}\dot{t}$, so that 
\begin{equation}
u_{\pm }^{\alpha }=K_{\pm }\left[ \frac{g(r)}{V(r)r^{2}},\pm \frac{\sqrt{g(r)%
}}{2r},\frac{a(r)}{V(r)r^{2}}\right]  \label{geodesics}
\end{equation}%
where the plus/minus signs refer to outgoing/ingoing geodesics. When $l=0$\
this can be solved explicitly, and we recover the results for geodesic
motion examined in ref. \cite{gimonhash}. The $K_{\pm }$'s are constants
related to the energy $E=\frac{K_{\pm }}{4}$. Choosing the normalization 
\begin{equation*}
g_{\alpha \beta }u_{+}^{\alpha }u_{-}^{\beta }=-1
\end{equation*}%
at some point $r=r_{0}$, we obtain 
\begin{equation*}
K_{+}K_{-}=\frac{2r_{0}^{2}V(r_{0})}{g(r_{0})}
\end{equation*}%
and for convenience we pick $K_{-}=1,K_{+}=\frac{2r_{0}^{2}V(r_{0})}{g(r_{0})%
}$. The expansion scalar for null geodesics is 
\begin{equation*}
\Theta (u_{\pm })=\pm \frac{K_{\pm }g^{\prime }(r)}{4r\sqrt{g(r)}}
\end{equation*}%
and we see that for outgoing null rays there is a sign difference between
geodesics starting inside the horizon ($r_{0}<r_{H})$ and geodesics starting
outside ($r_{0}>r_{H}$), with no such change for ingoing geodesics, as
expected for a trapped surface at $r=r_{H}$.

We can also say useful things about \textbf{\ }the CTC boundary. It occurs
when $g(r_{ctc})=0$, so the expansions are infinite there. Furthermore $%
\frac{dr}{dt}$ is infinite and $\frac{dr}{d\lambda }=0$ there, implying that
null geodesics cannot cross the CTC boundary. These results are consistent
with the observations for the Schwarzschild-G\"{o}del ($l=0$) case \cite%
{gimonhash}: null geodesics will take infinite coordinate time $t$ to go
between the black hole horizon and CTC boundary. The CTC boundary is reached
in finite affine parameter $\lambda $ although once the null ray reaches the
CTC horizon it spirals back toward the black hole.\textbf{\ }

\subsection{The Other Region of Parameter Space ($J^{2}>1$)}

When $x_{ctc}<x_{-}$, i.e. the CTC boundary is the innermost surface, it is
unclear what sort of object we now have.\textit{\ }For convenience we shall
continue to use the term horizon to signify $x_{\pm }$, and the term
ergosphere to denote the surface $x=1$\ ($r=\sqrt{2m}$), mindful of
potential abuses of language. Both horizons are now outside of the
ergosphere, but the CTC boundary can either be inside or outside of this
surface, depending on the choice of parameters. For example for $J=1.5,L=-2$%
, $x_{ctc}>1$\ (and $x_{ctc}<x_{-}<x_{+}$) but $x_{ctc}<1$\ for $J=2,L=-2$.

To understand the causal properties of this spacetime we shall consider the
metric for fixed $\theta $\ and $\psi $\ for convenience.\ \ Consider the
behavior of\textbf{\ }%
\begin{eqnarray}
ds^{2} &=&-\frac{r^{2}V(r)}{4g(r)}dt^{2}+g(r)(d\phi -\frac{a(r)}{g(r)}%
dt)^{2}+\frac{dr^{2}}{V(r)}  \notag \\
&=&\frac{r^{2}V(r)}{4\left| g(r)\right| }dt^{2}-\left| g(r)\right| (d\phi +%
\frac{a(r)}{\left| g(r)\right| }dt)^{2}+\frac{dr^{2}}{V(r)}
\label{simplified}
\end{eqnarray}%
where in the outer region $r>r_{+}$\ we see that $g(r)<0$\ and $V(r)>0$, and
so we have rewritten the metric for fixed $\left( \theta ,\psi \right) $\ in
the 2nd line above. For $r_{-}<r<r_{+}$\ then $g(r)<0$\ and $V(r)<0$.\ 

We can define a new coordinate $\chi =\phi -\frac{a(r_{0})}{g(r_{0})}t$\ \
for some $r_{0}>r_{+}$\ and the metric is now%
\begin{eqnarray}
ds^{2} &=&(-f(r)+g(r)\Omega ^{2}-2a(r)\Omega )dt^{2}+2(g(r)\Omega
-a(r))dtd\chi  \label{metricarbr} \\
&&+g(r)d\chi ^{2}+\frac{dr^{2}}{V(r)}  \notag
\end{eqnarray}%
where $\Omega =\frac{a(r_{0})}{g(r_{0})}$. For a fixed value of $%
r=r_{0}>r_{+}$\ this metric simplifies to:%
\begin{equation}
ds^{2}=+\frac{r_{0}^{2}V(r_{0})}{4\left\vert g(r_{0})\right\vert }%
dt^{2}-\left\vert g(r_{0})\right\vert d\chi ^{2}  \label{metricr0}
\end{equation}%
Since $g_{tt}>0,g_{\chi \chi }<0$\ we see that $\chi $\ functions as the
time coordinate, but only near $r=r_{0}$. \ For any given $r_{0}>r_{+}$ it
is possible to choose such a time coordinate $\chi $ in a neighbourhood of $%
r_{0}$\ and the metric always has signature $\left( -++++\right) $.

\ When $r_{-}<r<r_{+}$\ the signature of the metric becomes $\left(
---++\right) $\ \ and so this region is not a physical spacetime. \ There is
no choice of coordinate transformation that will allow the metric to have
correct signature. This is easily seen by expanding the metric near $r_{+}$%
\textbf{\ }\textit{\ \ }%
\begin{eqnarray*}
ds^{2} &=&+\frac{r_{+}^{2}V^{\prime }(r_{+})}{4\left\vert
g(r_{+})\right\vert }(r-r_{+})dt^{2}+2\left( g^{\prime }(r_{+})\Omega
_{+}-a^{\prime }(r_{+})\right) (r-r_{+})dtd\chi \\
&&-\left\vert g(r_{+})\right\vert d\chi ^{2}+\frac{dr^{2}}{V^{\prime
}(r_{+})(r-r_{+})}
\end{eqnarray*}%
indicating that the metric changes signature as $r$\ passes through $r_{+}$\
from above. \ There is a conical singularity at $r=r_{+}$ that is removed by
imposing a periodicity on $t$ of \ $\frac{r_{+}V^{\prime }(r_{+})}{8\pi 
\sqrt{|g(r_{+})|}}$. Hence the region $r>r_{+}$\ is a regular spacetime
everywhere permeated by closed timelike curves due to the periodicity of $%
\phi $ and $t$. \ 

\textit{\ }\ 
\begin{figure}[tbp]
\centering
\includegraphics[
height=3.1678in,
width=3.1678in
]{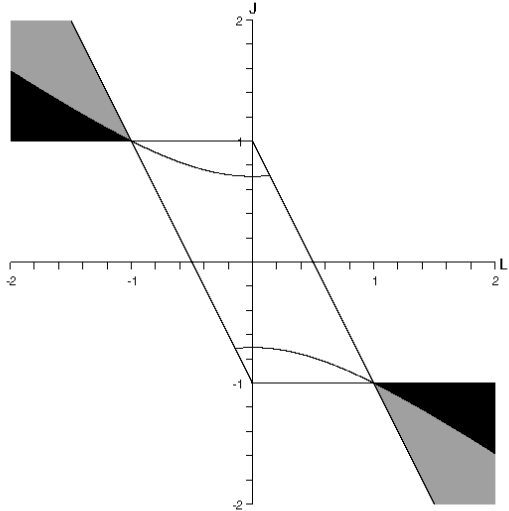}
\caption{Plot of of horizon behavior in terms of $J$ and $L$. \ Curve
corresponds to $r_{ctc}=\protect\sqrt{2m}$. \ In the black region $r_{ctc}>%
\protect\sqrt{2m}$ and in the grey region $r_{ctc}<\protect\sqrt{2m}.$}
\label{Figure3}
\end{figure}

Setting now $r<r_{-}$\ we need to consider two distinct cases depending on
where the \textquotedblleft ergosphere\textquotedblright \textit{\ }($%
r_{ergo}=$\ $\sqrt{2m}$) is located with respect to the CTC horizon. \ These
are $r_{ergo}<r_{ctc}<r_{-}$\ (represented by the black region in figure \ref%
{Figure3} in terms of $L$\ and $J$\ parameters) \ and $%
r_{ctc}<r_{ergo}<r_{-} $\ (represented by the grey region in figure \ref%
{Figure3} in terms of $L$\ and $J$\ parameters). \ \ 

We will start with the $r_{ergo}<r_{ctc}<r_{-}$\ case and we will restrict
ourselves to CTC region $r_{ctc}<r<r_{-}$ so that $g(r)<0$\ and $V(r)>0$. So
for this case the metric is once again in the same form as \ (\ref%
{simplified}). \ It is possible for an arbitrary $r_{0}$ ($%
r_{ctc}<r_{0}<r_{-}$) to choose a coordinate $\chi =\phi -\frac{a(r_{0})}{%
g(r_{0})}t$ and the metrics (\ref{metricarbr}) and (\ref{metricr0}) will be
valid in this region. \ Expanding the metric near $r_{-}$ gives 
\begin{eqnarray*}
ds^{2} &=&+\frac{r_{-}^{2}|V^{\prime }(r_{-})|}{4\left\vert
g(r_{-})\right\vert }(r_{-}-r)dt^{2}-2\left( g^{\prime }(r_{-})\Omega
_{-}-a^{\prime }(r_{-})\right) (r_{-}-r)dtd\chi \\
&&-\left\vert g(r_{-})\right\vert d\chi ^{2}+\frac{dr^{2}}{|V^{\prime
}(r_{-})|(r_{-}-r)}
\end{eqnarray*}%
again showing that when $r_{-}<r<r_{+}$\ the signature of the metric becomes 
$\left( ---++\right) $. Removal of the conical singularity at $r=r_{-}$\ is
achieved by imposing a periodicity on $t$\ of \ $\frac{r_{-}|V^{\prime
}(r_{-})|}{8\pi \sqrt{|g(r_{-})|}}$. \ Notice that this differs from that
imposed in the $r>r_{+}$\ region, as expected for two regions that are
disconnected spacetimes.\ \ Referring back to (\ref{metricarbr}), for an
arbitrary choice of $\chi $\ we see that $g_{\chi \chi }\rightarrow 0$ as
the CTC horizon is approached.\ At the CTC horizon $g_{tt}$\ can be either
positive or negative depending on the $\Omega $\ that defined $\chi $. \
Examining (\ref{metricarbr}) at $r=r_{ctc}$, it is clear that $g_{tt}$\ will
be positive if (a) $\Omega $\ has an opposite sign to $a(r_{ctc})$\ (in
general this will be true since $\Omega =\frac{a(r_{0})}{g(r_{0})}$ and $%
g(r_{0})$ is negative) and (b)$\ |\Omega |>\frac{f(r_{ctc})}{2|a(r_{ctc})|}$%
\ (i.e. $r_{0}$ must be chosen to be close enough to $r_{ctc}$ so that this
inequality will be satisfied); otherwise $g_{tt}$\ will be negative at the
CTC horizon. \ So for an arbitrary choice of parameters $L$\ and $J$\ it
will not be possible to choose a single coordinate $\chi $\ for which one
can write the metric in a form in which $\nabla t$\ is spacelike everywhere
between $r_{ctc}$\ and $r_{-}$. \ However for an arbitrary $r_{0}$\ we can
choose a coordinate $\chi $\ so that in a neighbourhood of $r_{0}$\ the
metric can be written with $\nabla t$\ spacelike. \textbf{\ }

\ The $r_{ctc}<r_{ergo}<r_{-}$\ case is a little more interesting due to the
presence of the ``ergosphere''. \ Outside of the ergosphere the analysis
remains the same as the previous $r<r_{-}$ case. \ Inside the ergosphere at
any given $r_{0}$ one can still choose $\chi =$\ $\phi -\frac{a(r_{0})}{%
g(r_{0})}t$ . However when $r_{ctc}<$\ $r<r_{ergo}$\ it is sufficient to
choose $\Omega =0$\ because $f(r)$ is negative and the metric 
\begin{eqnarray*}
ds^{2} &=&-f(r)dt^{2}-a(r)dtd\phi +g(r)d\phi ^{2}+\frac{dr^{2}}{V(r)} \\
&=&|f(r)|dt^{2}-a(r)dtd\phi +g(r)d\phi ^{2}+\frac{dr^{2}}{V(r)}
\end{eqnarray*}%
is such that $\phi $ is the time coordinate inside the ergosphere. \ 

\section{Temperature from Tunneling}

We now examine the application of the tunnelling method to the Kerr-G\"{o}%
del spacetime. \ For the case when the CTC horizon is outside the black hole
horizons the temperature has been previously computed by other means \cite%
{shwarz-godel thermo,rotating thermo}, allowing us to compare these with the
tunnelling results. We can also see if any tunnelling occurs from the CTC
horizon.

\subsection{Review of the Tunneling method}

The tunnelling method\ is a semi-classical approach that considers a
particle idealized as a spherical wave of matter\textbf{\ }emitted from
inside the horizon to outside. \ From the WKB approximation the tunneling
probability for the classically forbidden trajectory of the s-wave from
inside to outside the horizon is%
\begin{equation}
\Gamma \propto \exp (-2\mathrm{Im}I)
\end{equation}%
\ (here 
h{\hskip-.2em}\llap{\protect\rule[1.1ex]{.325em}{.1ex}}{\hskip.2em}
is set equal to unity). \ Expanding the action in terms of the particle
energy, the Hawking temperature is recovered at linear order. \ In other
words for $2I=\beta E+O(E^{2})$ this gives%
\begin{equation}
\Gamma \thicksim \exp (-2\func{Im}I)\simeq \exp (-\beta E)
\end{equation}

From this point there are two approaches that can be used to calculate the
imaginary part of the action, referred to as the null geodesic method and
the Hamilton-Jacobi Ansatz (refer to \cite{Last paper} for more details on
the two approaches). \ We will only use the null geodesic method in this
paper.\textbf{\ }

\ \ The imaginary part of the action for an outgoing s-wave (which follows a
radial null geodesic) from $r_{in}$ to $r_{out}$ is expressed as%
\begin{equation}
I=\int_{r_{in}}^{r_{out}}p_{r}dr=\int_{r_{in}}^{r_{out}}%
\int_{0}^{p_{r}}dp_{r}^{\prime }dr
\end{equation}%
where $r_{in}$ and $r_{out}$ are the respective initial and final radii of
the black hole. The trajectory between these two radii is the barrier the
particle must tunnel through. \ Note the local nature of this calculation:
the tunnelling probability only depends on an integration from $r_{in}$\ to $%
r_{out}$. \ In fact only the near horizon form of the black hole metric is
required in order to calculate the tunneling probability (and hence the
black hole temperature) \cite{Vargenas,Last paper}.\ So a stationary
observer anywhere outside the black hole would be able to observe the
emission and measure the temperature. \ This is in contrast with the Wick
Rotation method which requires a (scalar) field at infinity to be in
equilibrium with the black hole in order to get the temperature. \ The
presence of the CTC horizon at large distances renders the foundations of
this latter approach somewhat questionable.

We assume that the emitted s-wave has energy $\omega <<M$\ and that the
total energy of the space-time was originally $M$. Invoking conservation of
energy, to this approximation the s-wave moves in a background spacetime of
energy $M\rightarrow M-\omega ^{\prime }$. \ In order to evaluate the
integral, we employ Hamilton's equation $\dot{r}=\frac{dH}{dp_{r}}|_{r}$\ to
switch the integration variable from momentum to energy ($dp_{r}=\frac{dH}{%
\dot{r}}$), giving%
\begin{equation}
I=\int_{r_{in}}^{r_{out}}\int_{M}^{M-\omega }\frac{dr}{\dot{r}}%
dH=\int_{0}^{\omega }\int_{r_{in}}^{r_{out}}\frac{dr}{\dot{r}}(-d\omega
^{\prime })  \label{ssint}
\end{equation}%
where $dH=-d\omega ^{\prime }$\ because total energy $H=M-\omega ^{\prime }$%
\ with $M$\ constant.\ Note that $\dot{r}$\ is implicitly a function of $%
M-\omega ^{\prime }$. \ Since $\omega <<M$\ \ it is possible to rewrite the
expression in terms of an expansion of $\omega $. \ To first order this gives%
\textbf{\ \ }%
\begin{align}
I& =\int_{0}^{\omega }\int_{r_{in}}^{r_{out}}\frac{dr}{\dot{r}(r,M-\omega
^{\prime })}(-d\omega ^{\prime })=-\omega \int_{r_{in}}^{r_{out}}\frac{dr}{%
\dot{r}(r,M)}+O(\omega ^{2})  \notag \\
& \simeq \omega \int_{r_{out}}^{r_{in}}\frac{dr}{\dot{r}(r,M)}
\label{nullaction}
\end{align}%
To proceed further we will need to estimate the last integral. First we note
that $r_{in}>r_{out}$\ because black holes decrease in mass as energy is
emitted; consequently the radius of the event horizon decreases. \ We
therefore write $r_{in}=r_{H}(M)-\epsilon $\ and $r_{out}=r_{H}(M-\omega
)+\epsilon $\ where $r_{H}(M)$\ denotes the location of the event horizon of
the original background space-time before the emission of particles. \
Henceforth the notation $r_{H}$\ will be used to denote $r_{H}(M)$. Note
that with this generalization no explicit knowledge of the total energy or
mass is required since $r_{H}$\ is simply the radius of the event horizon
before any particles are emitted.

We pause to discuss a few technical points connected with rotating
spacetimes \cite{Vargenas},\cite{Last paper}. \ In general the emitted
s-wave could carry angular momentum $\ell $; if it has energy $E$\ then the
tunnelling probability to the lowest order would be 
\begin{equation*}
\Gamma \simeq \exp (-\beta (E-\Omega _{H}\ell ))
\end{equation*}%
where $\Omega _{H}$\ is the angular velocity of the black hole horizon. \
For this tunnelling probability to make sense we must require $E-\Omega
_{H}\ell >0$. \ This inequality corresponds to the s-wave being able to
escape from the ergosphere. For calculating the temperature $\beta ^{-1}$\
it is sufficient to restrict to $\ell =0$\ s-waves.

\subsection{Temperature Calculation}

Turning now to calculation of the black hole temperature \cite{early
tunnelling,Recent Tunnelling}, recall that\ the full metric in lapse shift
form is (\ref{LSform}). \ To employ the null geodesic method it is
convenient to write the metric in a Painlev\'{e} form so that the null
geodesic equations convey the semi-permeable nature of the black hole
horizon (i.e. that it is easy to cross into the black hole but classically
they cannot escape).\textit{\ }\ In order to simplify the equations we
rewrite the metric by defining $\chi =\phi -\Omega _{H}t$\textbf{. }We are
interested in geodesics that have no angular momentum ($\ell =0$) so we set $%
d\chi =0$\ (and for convenience also $d\theta =d\psi =0$), yielding

\begin{equation*}
ds^{2}=-\frac{r^{2}V(r)}{4g(r)}dt^{2}+\frac{dr^{2}}{V(r)}
\end{equation*}

We can easily rewrite this in Painlev\'{e} form via the following
transformation%
\begin{equation*}
t\rightarrow t-\dint \frac{2\sqrt{g(r)}}{rV(r)}\sqrt{1-V(r)}dr
\end{equation*}%
giving (for constant $\chi ,\theta ,$and $\psi $) the following Painleve
metric 
\begin{equation*}
ds^{2}=-\frac{r^{2}V(r)}{4g(r)}dt^{2}+\frac{r}{\sqrt{g(r)}}\sqrt{1-V(r)}%
drdt+dr^{2}
\end{equation*}%
\textit{\ }

We need to know how the $\ell =0$ null geodesics behave for this metric in
order to solve for the imaginary part of the action using equation (\ref%
{nullaction}). \ The radial null geodesic equation is given by:%
\begin{equation}
\frac{dr}{dt}=\frac{r}{2\sqrt{g(r)}}(\pm 1-\sqrt{1-V(r)})
\label{radial null}
\end{equation}%
where $+$\ denotes outgoing and $-$\ denotes ingoing geodesics (notice that $%
\frac{dr}{dt}=0$\ at the horizon for outgoing geodesics and $\frac{dr}{dt}$\
is nonzero for ingoing geodesics). \ Inserting (\ref{radial null}) into (\ref%
{nullaction}) we find that $1/\frac{dr}{dt}$has a first order pole at the
horizon with residue $\frac{4\sqrt{g(r_{H})}}{r_{H}V^{\prime }(r_{H})}$. \
So solving the integral we find:%
\begin{eqnarray*}
\func{Im}I &=&\frac{4\pi \omega \sqrt{g(r_{H})}}{r_{H}V^{\prime }(r_{H})}%
+O(\omega ^{2}) \\
\Gamma &\thicksim &\exp (-2\func{Im}I)\simeq \exp (-\frac{8\pi \sqrt{g(r_{H})%
}}{r_{H}V^{\prime }(r_{H})}\omega )
\end{eqnarray*}

This corresponds to a temperature: \ 

\begin{eqnarray}
T &=&\frac{r_{H}V^{\prime }(r_{H})}{8\pi \sqrt{g(r_{H})}}
\label{general temp} \\
T &=&\frac{m(r_{H}^{2}(1-8j^{2}m-4ml)-2l^{2})}{\pi r_{H}^{3}\sqrt{%
-4j^{2}r_{H}^{6}+(1-8j^{2}m)r_{H}^{4}+2ml^{2}}}  \label{specific Temp}
\end{eqnarray}%
This temperature is the same as that obtained using Wick-rotation methods 
\cite{rotating thermo,general thermo}; when $l=0$\ it reduces down to the
Schwarzschild-G\"{o}del temperature found in \cite{shwarz-godel thermo}.
Note that the expression for the temperature diverges when $g(r_{H})=0$,
which occurs when the CTC\ horizon is coincident with the outer horizon. \
The temperature is not defined when $g(r_{H})<0$\ , an unsurprising result
considering the analysis of the other region of parameter space and the fact
the when the CTC horizon is inside the $r_{-}$\ and $r_{+}$\ horizons the
derivation used is not valid. \ Not only is $t$\ not the correct time
coordinate, but it is unclear how to even define tunnelling from inside $%
r_{+}$\ because the region $r_{-}<r<r_{+}$\ is not a spacetime.\textit{\ \ }

Consider next what happens if we try to apply the tunnelling method to the
CTC horizon. \ From (\ref{radial null}) we know that $\frac{dr}{dt}%
\rightarrow \infty $\ as $r\rightarrow r_{ctc}$. \ This means that $1/\frac{%
dr}{dt}$\ is simply zero at the CTC horizon. Since $1/\frac{dr}{dt}$\ has no
poles at the CTC horizon it means there is no tunnelling at the CTC\
horizon. \ \ 

\section{Conclusions}

In this paper we have reviewed some of the general properties of the Kerr-God%
\"{e}l spacetime, performing\ detailed analysis of its parameter space.
There are two distinct classes. One is the class $J^{2}<1$, corresponding to
black holes for which the CTC horizon $r_{CTC}$\ is exterior to the black
hole horizons at $r_{+}$\ and $r_{-}$\ . When $J^{2}>1$\ we obtain the other
class (the ``other''\ region of parameter space), for which the CTC horizon
is inside both of the other surfaces $r_{+}$\ and $r_{-}$\ . \ We find that
these are the only two possibilities (apart from naked singularities); there
is no ``in-between''\ region where $r_{+}>r_{CTC}>r_{-}$\ , contrary to
previous expectations \cite{general thermo}. \ 

Despite the presence of CTCs, we find that the tunnelling method applied to
the black hole region of parameter space yields a temperature consistent
with previous calculations made via Wick rotation methods. \ We also find
(when $r_{ctc}>r_{+}$\ ) that there is no tunnelling through the CTC
horizon. \ We have discussed technical problems that occur in trying to
apply the tunnelling method to the ``other''\ region of parameter space due
the fact that the region $r_{-}<r<r_{+}$\ does not have the correct
signature. \ Higher-order corrections and applications of the method to
non-radial null rays remain as interesting problems to explore.

{\Huge \bigskip Acknowledgements}

This work was supported in part by the Natural Sciences and Engineering
Research Council of Canada.

\end{document}